\documentclass[pre,twocolumn,amsmath,amssymb,showpacs]{revtex4}

\usepackage{graphicx}
\usepackage{amssymb}
\usepackage{epstopdf}
\usepackage{times}
\usepackage{dcolumn}
\usepackage{bm}
\usepackage{amsmath}
\usepackage[dvips]{color}

\newcommand{\bfn}{\mathbf{n}} 
\newcommand{\bfPsi}{\mathbf{\Psi}}  
 
\newcommand{\caE}{\mathcal{E}}     
\newcommand{\caF}{\mathcal{F}}   
\newcommand{\caK}{\mathcal{K}}    
\newcommand{\caH}{\mathcal{H}}   
\newcommand{\caN}{\mathcal{N}}      
\newcommand{\caX}{\mathcal{X}}  
\newcommand{\dL}{\delta l}     
     
\newcommand{\tN}{N} 
\newcommand{\spaEq}{\hspace{0.5cm}}

\newcommand{\stateR}[1]{\left|#1\right\rangle} 
\newcommand{\stateL}[1]{\left\langle #1\right|} 
\newcommand{\sumindex}[2]{\substack{#1 \\ (#2)}}

\newcommand{\vspfigA}{\vspace{0.0cm}}  
\newcommand{\vspfigB}{\vspace{0.0cm}} 
\newcommand{\vspfigC}{\vspace{0.2cm}}
\newcommand{\widthfig}{0.4\textwidth}

\begin{document}



\title{Escape Behavior of Quantum Two-Particle Systems with Coulomb Interactions}

\author{Tooru Taniguchi and Shin-ichi Sawada}

%
\affiliation{School of Science and Technology, Kwansei Gakuin University, 
2-1 Gakuen, Sanda city, Hyogo, Japan} 

\date{\today}
\begin{abstract}

   Quantum escapes of two particles with Coulomb interactions from a confined one-dimensional region to a semi-infinite lead are discussed by the probability of particles remaining in the confined region, i.e. the survival probability, in comparison with one or two free particles. 
   For free-particle systems the survival probability decays asymptotically in power as a function of time. 
   On the other hand, for two-particle systems with Coulomb interactions it shows an exponential decay in time. 
   A difference of escape behaviors between Bosons and Fermions is considered as quantum effects of identical two particles such as the Pauli exclusion principle. 
   The exponential decay in the survival probability of interacting two particles is also discussed in a viewpoint of quantum chaos based on a distribution of energy level spacings. 
   
\end{abstract}
   
   

\maketitle  
\section{Introduction}

   The escape is a behavior of open systems in which materials move out from an observed area. 
   It has drawn considerable attention in various points of view, for example, Kramers' escape problem \cite{R89,K92,HTB90}, $\alpha$-decaying nucleus \cite{GC29,W61,DN02}, the first-passage time problem \cite{R89,K92,HTB90}, the recurrence time problem \cite{AT09}, the controlling chaos \cite{PB00}, and the Riemann hypothesis \cite{BD05}, etc. 
   Escapes involve transport, and can be used to calculate transport coefficients \cite{GN90,G98,K07}. 
   Particles escaping from thermal reservoirs can sustain flows such as electric currents \cite{D95,I97}. 
   Escape phenomena have been investigated in many different systems, e.g. billiard systems (by theories \cite{BB90,AGH96,BD07} and by experiments \cite{MHC01,FKC01}), map systems \cite{AT09,PB00,DY06}, wave dynamics \cite{RLK06,SH07}, and stochastic systems \cite{R89,K92,HTB90}, etc. 

   A typical quantity to characterize a particle escape is the probability of particles remaining within the observed area from which particles can move out, the so-called survival probability. 
   The survival probability decays in time because particles keep to escape from the observed area without coming back, and its decay properties have been an important subject in chaotic dynamics \cite{G98,K07,TG06}. 
   In classical billiard systems, it is conjectured, based on an ergodic argument, that the survival probability decays exponentially for chaotic systems, while it shows a power decay for non-chaotic systems \cite{BB90}. 
   This conjecture has been examined in detail, e.g. in a finite size effect of holes \cite{AGH96}, weakness of chaos \cite{AT09}, a connection to correlation functions \cite{BD07}, and a deviation from an escape rate estimated by the natural invariant measure \cite{AT09,PB00}. 
   
   Particle escapes have been also discussed in quantum systems by using the survival probability. 
   Ref. \cite{M03} discussed an escape behavior of a free particle in a one-dimensional system by a concrete calculation of wave-packet dynamics, and Ref. \cite{DN02} investigated escapes of a particle with a potential barrier.    
   Quantum escapes have been also considered by using random matrix approach for quantum scattering systems \cite{LW91,DHM92}, numerical approaches to wave-packet dynamics in quantum billiard systems \cite{ZB03}, in a viewpoint of chaotic dynamics. 
   These studies show that the survival probability decays in power or exponentially, depending on how quantum states are superposed initially, but a general quantum mechanical decay behavior of the survival probability based on chaotic features has not been clearly established yet.   

   The principal aim of this paper is to investigate many-particle effects in escape behaviors of quantum chaotic systems in comparison with non-chaotic free-particle systems. 
   To investigate them in systems as simple as possible, we consider particle escapes from a confined one-dimensional region to a semi-infinite one-dimensional lead.  
   Furthermore, as a simple chaotic many-particle system we choose the system consisting of two particles with Coulomb interactions, and discuss many-particle effects by comparing the chaotic two-particle cases with the non-chaotic cases of one or two free particles.
   In these situations we consider particle escapes whose initial states are represented as an energy eigenstate of particles confined spatially at the initial time, so that voluntariness of initial superposition of quantum states in the escape dynamics does not appear in discussions of this paper. 
   We calculate survival probabilities of such systems as a function of time, and show that the survival probability of the Coulomb-interacting two-particle system decays exponentially in time, while for the free-particle systems it decays in power as a function of time. 
   We also discuss quantum effects of identity of two particles, like the Pauli exclusion principle, in quantum escape problems, appearing as a difference of escape behaviors between Bosons and Fermions. 
   It is also shown that a confined system consisting of two particles with Coulomb interactions has a repulsive feature of energy level spacings, which is regarded as a character of quantum chaos. 
   Noting that one-dimensional one-particle systems, as well as two free particles in a one-dimensional space, cannot be chaotic, this may imply a relation of the exponential decay of survival probability with quantum chaos.

\section{Escape of Many Particles in a Semi-Infinite One-Dimensional Space}

\begin{figure}[!t]
\vspfigA
\begin{center}
\includegraphics[width=\widthfig]{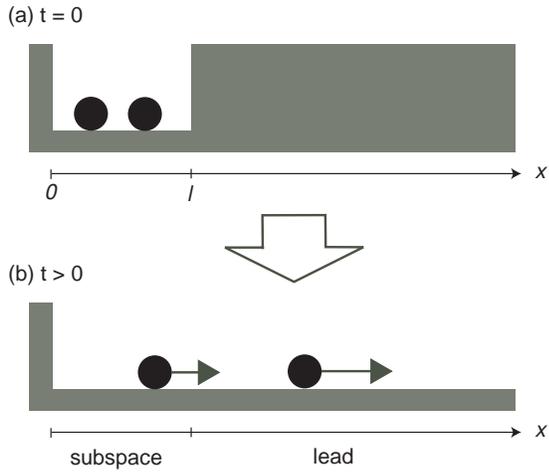}
\vspfigB
\caption{
   Schematic illustration of a particle escape in a semi-infinite one-dimensional system. 
   (a) Particles confined inside a finite region $[0,l]$ at $t =0$. 
   (b) Particles escaping to a semi-infinite region at $t>0$, by removing the infinite potential barrier in the region $[l,+\infty)$. 
   In this situation we call the region $[0,l]$ as the subspace, and the region $(l,+\infty)$ as the lead. 
}
\label{fig1System}
\end{center}
\vspfigC
\end{figure}

   In this paper, we consider quantum systems consisting of $\tN$ particles in a one-dimensional semi-infinite region $[0,+\infty)$. 
   Before the initial time $t < 0$, we set the infinite potential barrier in the region $[l,+\infty)$, and confine the particles in the finite region $[0,l]$ with a positive constant $l$. 
   At the initial time $t=0$ we remove this infinite potential barrier in $x\geq l$, so that a particle escape to the region  $[l,+\infty)$ becomes to occur. 
   The schematic illustration of this escaping behavior is shown in Fig. \ref{fig1System}. 
   (Here, the particles in Fig. \ref{fig1System} are drawn as particles with a nonzero finite size to make them visible, but in the actual models used in this paper we consider material particles with a infinitesimally small size.)
   To make a clear image of this kind of escape phenomena, we call the region $[0,l]$ as the ``subspace'', and the region $(l,+\infty)$ as the ``lead,'' so the particle escape occurs from the subspace to the lead.  
%

   This system is described by the wave function $\Psi(x_{1},x_{2},\cdots,x_{\tN},t)$ at time $t$ as a solution of the Schr\"odinger equation 
\begin{eqnarray}
   i\hbar \frac{\partial \Psi(x_{1},x_{2},\cdots,x_{\tN},t)}{\partial t} 
   =  \hat{H}  \Psi(x_{1},x_{2},\cdots,x_{\tN},t), 
\label{SchroEquat1}
\end{eqnarray}
where  $\hat{H}$ is the Hamiltonian operator, $\hbar$ is the Dirac constant $\hbar$, and $x_{j}$ is the position of the $j$-th particle, $j=1,2,\cdots,\tN$. 
   If the system consists of identical particles in the quantum mechanical sense, then the wave function $\Psi(x_{1},x_{2},\cdots,x_{\tN},t)$ must satisfy 
\begin{eqnarray}
   && \Psi(x_{1},x_{2},\cdots,x_{\tN},t) 
      \nonumber \\
   &&\spaEq = \pm \Psi(x_{1},x_{2},\cdots,x_{\tN},t)|_{(x_{j},x_{k}) 
   \rightarrow (x_{k},x_{j})}
\label{PauliExclu1}
\end{eqnarray}
for exchange of any particle indices $j$ and $k$, $j=1,2,\cdots,\tN$, $k=1,2,\cdots,\tN$. 
   Here, the sign $-$ ($+$) in $\pm$ of the right-hand side of Eq. (\ref{PauliExclu1}) should be taken for Fermions (Bosons) \cite{Note1}, imposing the Pauli exclusion principle for Fermions. 
   Using the wave function $\Psi(x_{1},x_{2},\cdots,x_{\tN},t)$ satisfying Eq. (\ref{SchroEquat1}) we introduce the probability $P(t)$ defined by  
\begin{eqnarray}
   && P(t) \equiv \int_{0}^{l} dx_{1}\int_{0}^{l} dx_{2} \cdots\int_{0}^{l} dx_{\tN}\; 
      \nonumber \\
   &&\spaEq\spaEq\spaEq\spaEq\times 
      |\Psi(x_{1},x_{2},\cdots,x_{\tN},t)|^{2} .
\label{SurviProba1}
\end{eqnarray}
   This is the probability with which $\tN$ particles are still inside the subspace and have not escaped to the lead at time $t$ yet, and we call it the ``survival probability'' hereafter in this paper  \cite{Note2}.  
   The purpose of this paper is to discuss an escape behavior of particles 
by a time-dependence of the survival probability $P(t)$. 
   It may be noted that the subspace can be regarded as an open system coupled to a semi-infinite lead, but the phenomena considered here are not scattering phenomena described by  
a response of the system to incident waves, in the sense that particles are always exist only in a finite region at any finite time and the wave function is normalizable, i.e. $\int_{0}^{+\infty}dx_{1}\int_{0}^{+\infty}dx_{2}\cdots\int_{0}^{+\infty} dx_{\tN}\; |\Psi(x_{1},x_{2},\cdots,x_{\tN},t)|^{2} =1$ at any time $t$, different from scattering states including an incoming plain wave from the infinite region.  

   In general, the survival probability $P(t)$ depends on the initial condition. 
   As an initial condition at the time $t=0$, in this paper we choose an energy eigenstate $\Phi_{n}(x_{1},x_{2},\cdots,x_{\tN})$ of $N$ particles confined inside the subspace region $[0,l]$ corresponding to energy eigenvalue $E_{n}$ $(E_{1}\leq E_{2} \leq \cdots )$. 
   In this way, we obtain the survival probability for each initial wave function  $\Psi(x_{1},x_{2},\cdots,x_{\tN},0) = \Phi_{n}(x_{1},x_{2},\cdots,x_{\tN})$, and present it as $P_{n}(t)$, $n=1,2,\cdots$.   

   Under such initial conditions, we calculate the survival probability $P_{n}(t)$ analytically for free particle cases in Secs. \ref{EscapeOneParticle} and \ref{TwoFreeParticles} in this paper. 
   We also show numerical results of the survival probability by discretizing space and time in the Schr\"odinger equation for two-particle systems in Sec. \ref{CoulombInteractionEscape}, as well as for a part of one-particle case in Sec. \ref{EscapeOneParticle}.  
   As a numerical technique we use the pseudo-spectral method \cite{FF82,D94}.  
   As an example, in Appendix \ref{SpaceTimeDiscrite} we outline a spatial discretization of the Schr\"odinger equation for two-particle cases and its time-discretization by the pseudo-spectral method, which are used to calculate the numerical results shown in this paper.
   In these numerical calculations we take the unit of $m=1$ for the particle mass, $l=1$ for the length of the subsystem, and $\hbar =1$ for the Dirac constant. 
   For numerical calculations the one-dimensional space is discretized by the length $\delta x = l/\caN_{0}$ with the integer site number $\caN_{0}$ of the subspace (See Appendix \ref{SpaceDiscrite}.). 
   For numerical calculations by pseudo-spectral method, we also discretize the time by $\delta t$ (See Appendix \ref{PseudoSpectralMethod}.), and choose the concrete value of $\delta t$ so that the average energy and the normalization of wave function are almost conserved during the numerical calculations. 
   The total system length consisting of the subsystem and the lead in our numerical calculations is chosen as $L=\caN \delta x = \caN l/\caN_{0}$ with the total site number $\caN$ (so that the site number of the lead is given by $\caN-\caN_{0}$), and we calculate the particle escape dynamics in a time interval in which particle's returning back to the subsystem from the lead region is negligible.   
   Concrete values of the parameters $\caN_{0}$, $\caN$, and $\delta t$ will be shown for each numerical result in this paper.

\section{Escape of One Free Particle}
\label{EscapeOneParticle}

   We first consider the case of a single free particle in a semi-infinite one-dimensional space, whose Hamiltonian is given by 
$
   \hat{H} = -[\hbar^{2}/(2m)] \partial^{2}/\partial x^{2}
$
with the particle position $x \equiv x_{1}$. 
   Using this Hamiltonian we solve the Schr\"odinger equation $i\hbar \partial \Psi(x,t) /\partial t =  \hat{H}  \Psi(x,t)$ for the wave function $\Psi(x,t)$ of the system for $x\geq 0$ and $\Psi(0,t) = 0$, and calculate the survival probability (\ref{SurviProba1}). 
   
   In this system, as shown in Appendix \ref{EscapeFreeParticle}, we can solve the Schr\"odinger equation analytically, and the survival probability $P(t)$ is represented asymptotically as 
\begin{eqnarray}
   P(t) \;\overset{t\rightarrow+\infty}{\sim}\; \frac{A_{1}}{t^{3}} .
\label{EscapRateFree1}
\end{eqnarray}
   Here, the constant $A_{1}$ is given by 
\begin{eqnarray}
   A_{1}\equiv \frac{2}{3\pi} \left(\frac{ml}{\hbar}\right)^{3}\left|\int_{0}^{l}dx\;  
   x \; \Psi(x,0)\right|^{2}
\label{ContA1}
\end{eqnarray}
with the wave function $\Psi(x,0)$ at the initial time $t=0$.
   Equation (\ref{EscapRateFree1}) means that the survival probability $P(t)$ decays by power $\sim t^{-3}$ asymptotically in time, for arbitrary initial conditions of the subspace as far as $A_{1}\neq 0$. 
   Power decays of the survival probability for one-particle systems in a one-dimensional space have been discussed in some papers \cite{DN02,M03}. 

\begin{figure}[!t]
\vspfigA
\begin{center}
\includegraphics[width=\widthfig]{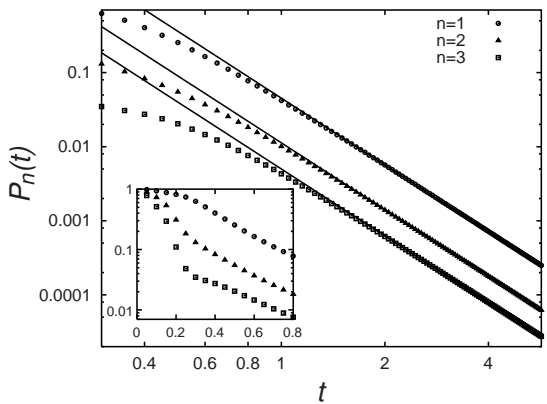}
\vspfigB
\caption{
   Survival probabilities $P_{n}(t), n=1,2,3$ of one free particle in a semi-infinite one-dimensional space as a function of time $t$ (the main figure as log-log plots for a long time behavior and the inset as linear-log plots for a short time behavior), corresponding to the energies $E_{1}$ (circle), $E_{2}$ (triangle), and $E_{3}$ (square), respectively. 
   The lines are plots of Eq. (\ref{EscapRateFree1}) with the coefficient (\ref{ContA1}) for the cases of energy $E_{n}$, $n=1,2,3$, which are proportional to $t^{-3}$.
}
\label{fig2OnePartiEsc}
\end{center}
\vspfigC
\end{figure}  
%
   Figure \ref{fig2OnePartiEsc} is the graphs of the survival probabilities $P_{n}(t)$, $n=1,2,3$ obtained by solving the Schr\"odinger equation for the one free particle numerically, using the subspace site number $\caN_{0}=60$, the total space site number $\caN = 32768$, and the discretized time interval $\delta t = 10^{-2}$.
   The energy values corresponding to these graphs are $E_{1} = 4.77$, $E_{2}=  19.1$, and $E_{3}= 42.9$.    
   For a comparison, in this figure we also draw the graph of Eq. (\ref{EscapRateFree1}) with the coefficient (\ref{ContA1}) for each energy.  
   The numerical results of the survival probabilities $P_{n}(t), n=1,2,3$ in Fig. \ref{fig2OnePartiEsc} show a behavior of the power decay $\sim t^{-3}$ independent of $n$ in a large time region, and they are consistent with our analytical result (\ref{EscapRateFree1}) including value of the coefficient (\ref{ContA1}). 

\begin{figure}[!t]
\vspfigA
\begin{center}
\includegraphics[width=\widthfig]{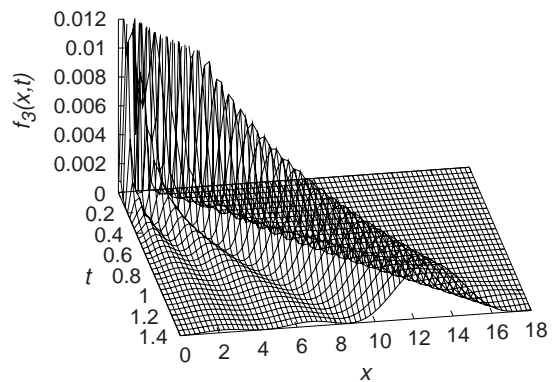}
\vspfigB
\caption{
   Probability distribution function $f_{3}(x,t) \equiv |\Psi_{3}(x,t)|^{2}$ of the particle position as a function of $t$ and $x$, corresponding to the energy $E_{3}$, for one free particle in a semi-infinite one-dimensional space $x\geq 0$. 
}
\label{fig2OneSpaProb}
\end{center}
\vspfigC
\end{figure}  
%
   In order to visualize an escape behavior of one-free-particle systems, we show in Fig. \ref{fig2OneSpaProb} the probability distribution function $f_{n}(x,t) \equiv |\Psi_{n}(x,t)|^{2}$ for $n=3$ as a function of time $t$ and position $x$.  
   Here, we took the case of $n=3$ in Fig. \ref{fig2OneSpaProb} so that this spatial distribution $f_{3}(x,t)$ has three peaks at the initial time $t=0$, but two of these three peaks decay quickly in time and only one peak survives for a long time and moves away from the subspace region $[0,l]$.

\section{Escape of Identical Two Free Particles}
\label{TwoFreeParticles}

   As a many-particle effect in escape phenomena, we first discuss the quantum effect of identity of two particles, such as the Pauli exclusion principle, in a semi-infinite one-dimensional system. 
   The Hamiltonian operator of the system is given by 
$\hat{H} = -[\hbar^{2}/(2m)] \left(\partial^{2}/\partial x_{1}^{2}+\partial^{2}/\partial x_{2}^{2}\right) $
with the position  $x_{j}$ of the $j$-th particle ($j=1,2$), and we impose the condition (\ref{PauliExclu1}) for the wave function $\Psi(x_{1},x_{2},t)$ as 
\begin{eqnarray}
   \Psi(x_{1},x_{2},t) = \pm \Psi(x_{2},x_{1},t) .
\label{ExchaInter1}
\end{eqnarray}
   Here, the sign $+$ ($-$) in the right-hand side of Eq. (\ref{ExchaInter1}) is taken when the particles are identical Bosons (Fermions).  
   The condition (\ref{ExchaInter1}) is guaranteed at any time $t$ as far as it is imposed at the initial time $t=0$ because the Hamiltonian operator $\hat{H}$ is symmetric for exchange of two positions of the particles.

\subsection{Boson Case}

   In the case of identical two free Bosons, as shown in Appendix \ref{TwoFreeBosons}, the asymptotic behavior of the survival probability $P(t)$ is represented as 
\begin{eqnarray}
   P(t) \;\overset{t\rightarrow+\infty}{\sim}\; \frac{A_{2b}}{t^{6}} .
\label{EscFreeTwoBoson1}
\end{eqnarray}
   Here, $A_{2b}$ is defined by 
\begin{eqnarray}
   A_{2b}&\equiv& \frac{4}{9\pi^{2}}\left(\frac{ml}{\hbar}\right)^{6}
   \left|\int_{0}^{l}dx_{1}\int_{0}^{l}dx_{2}\; x_{1}x_{2}\Psi(x_{1},x_{2},0)\right|^{2}
   \nonumber \\
\label{ConstBoson1}
\end{eqnarray}
with the initial wave function $\Psi(x_{1},x_{2},0)$. 
   Therefore, in the case of identical two free Bosons, the survival probability $P(t)$ asymptotically  decays in power $\sim t^{-6}$, which is simply the square of the result of the one free particle discussed in Sec. \ref{EscapeOneParticle}.

\subsection{Fermion Case}
\label{PauliExclusion}

   In the case of identical two free Fermions, as shown in Appendix \ref{TwoFreeFermions}, the asymptotic behavior of the survival probability $P(t)$ is represented as 
\begin{eqnarray}
   P(t) \;\overset{t\rightarrow+\infty}{\sim}\; \frac{A_{2f}}{t^{10}}
\label{EscFreeTwoFermi1}
\end{eqnarray}
with the constant $A_{2f}$ defined by 
%
\begin{eqnarray}
   A_{2f}&\equiv& \frac{2}{4725 \pi^{2}}\left(\frac{ml}{\hbar}\right)^{10} 
      \Biggl|\int_{0}^{l}dx_{1}\int_{0}^{l}dx_{2} \;
      \nonumber \\
   &&\spaEq\spaEq\spaEq\times 
      x_{1}x_{2}\left(x_{1}^{2}-x_{2}^{2}\right)
      \Psi(x_{1},x_{2},0) \Biggr|^{2} . 
      \spaEq
\label{ContFermi1}
\end{eqnarray}
%
   It is important to note that in the case of identical two free Fermions, the survival probability $P(t)$ asymptotically decays in the power $\sim t^{-10}$, which is different from the power  $\sim t^{-6}$ in the corresponding Boson case shown in Eq. (\ref{EscFreeTwoBoson1}). 
   In the identical two free Fermions the power decay term $\sim t^{-6}$ of the survival probability disappears by an cancellation for the Pauli exclusion principle $\Phi(x_{1},x_{2},t) + \Phi(x_{2},x_{1},t) = 0$, and such Fermions escape faster qualitatively than the corresponding identical two free  Bosons.

\section{Escape of Two Particles with Coulomb interactions}
\label{CoulombInteractionEscape}

   Now, we discuss an escape behavior of a semi-infinite-one-dimensional system consisting of two particles with Coulomb interactions. 
   The Hamiltonian operator of the system is given by 
$\hat{H} = -[\hbar^{2}/(2m)] \left(\partial^{2}/\partial x_{1}^{2}+\partial^{2}/\partial x_{2}^{2}\right) + U(x_{1},x_{2})$ where $U(x_{1},x_{2}) $ is the Coulomb potential energy as 
\begin{eqnarray}
   U(x_{1},x_{2}) = \frac{\lambda}{\sqrt{d^{2}+(x_{1}-x_{2})^{2}}} .
\label{Poten1}
\end{eqnarray}
   Here, $\lambda$ is given by $\lambda = q^{2}/(4\pi\epsilon_{0})$ with the particle charge $q$ and the dielectric constant $\epsilon_{0}$, and $d$ is a small but nonzero constant appearing as an effect of a quasi-one-dimensionality of the system \cite{FST01}. 
   Using this Hamiltonian we solve the Schr\"odinger equation $i\hbar \partial \Psi(x_{1},x_{2},t) /\partial t =  \hat{H}  \Psi(x_{1},x_{2},t)$ for the wave function $\Psi(x_{1},x_{2},t)$ of this system with the condition (\ref{ExchaInter1}), then calculate the survival probability (\ref{SurviProba1}). 
   
   In this section, we show only results of the identical-two-Fermion case, since the corresponding results for the Boson case are quite similar to the Fermion case and we could not find a particular difference (e.g. in an exponential decay behavior of the survival probability and value of its escape rate as will be shown for Fermions in this section) between the Fermion cases and the corresponding Boson cases in our numerical results.

\begin{figure}[!t]
\vspfigA
\begin{center}
\includegraphics[width=\widthfig]{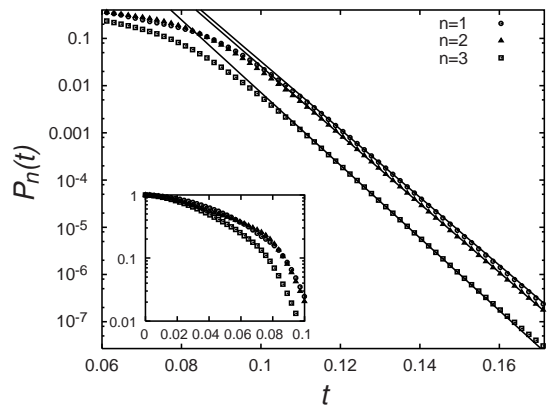}
\vspfigB
\caption{
   Survival probabilities $P_{n}(t), n=1,2,3$ for two particles with Coulomb interactions in a semi-infinite one-dimensional space as a function of time $t$ (linear-log plots), corresponding to the energies $E_{1}$ (circle), $E_{2}$ (triangle), and $E_{3}$ (square), respectively. 
   The inset is linear-log plots of $P_{n}(t)$ for these energies in a short time region. 
   The lines are the fits for each survival probability $P_{n}(t)$ to an exponential function $\alpha \exp(-\beta t)$ with fitting parameters $\alpha$ and $\beta$.
}
\label{fig3TwoPartiEsc}
\end{center}
\vspfigC
\end{figure}  
%
   Figure \ref{fig3TwoPartiEsc} is the survival probabilities $P_{n}(t)$, $n=1,2,3$ at time $t$ for identical two Fermions with Coulomb interactions in a semi-infinite one-dimensional space. 
   For numerical calculations to obtain the graphs in Fig. \ref{fig3TwoPartiEsc}, we used the parameter values $\lambda = 60$ and $d=10^{-4}/6$ for the potential energy (\ref{Poten1}), and also the subspace site number $\caN_{0}=60$, the total one-dimensional space site number $\caN = 2048$ (so the total site number in two-dimensional $x_{1}x_{2}$ space is $\caN^{2}=4194304$), and the discretized time interval $\delta t = 10^{-3}$. 
   The energy values corresponding to the survival probabilities $P_{n}(t), n=1,2,3$ in Fig. \ref{fig3TwoPartiEsc} are given by $E_{1} = 149$, $E_{2} =  202$, and $E_{3} = 213$, respectively. 
   By the repulsive Coulomb interaction, the survival probability $P_{n}(t)$ decays much more rapidly than the case of one free particle. 
   Figure \ref{fig3TwoPartiEsc} shows that the decay of the survival probability $P_{n}(t)$ for Coulomb-interacting two particles is well approximated as an exponential decay after a short time, different from the cases of one particle discussed in Sec. \ref{EscapeOneParticle}. 
   In Fig. \ref{fig3TwoPartiEsc} we also showed fits for each survival probability $P_{n}(t)$ to an exponential function $\alpha \exp(-\beta t)$ with the fitting parameter $\alpha$ and $\beta$. 
   Here, the values of fitting parameters are chosen as   
$(\alpha,\beta) = (5.03\times 10^{5},166)$ for  $P_{1}(t)$, 
$(\alpha,\beta) = (4.81\times 10^{5},167)$ for  $P_{2}(t)$, and 
$(\alpha,\beta) = (3.21\times 10^{5},176)$ for  $P_{3}(t)$. 
   The escape rate, defined as the parameter $\beta$, does not depend strongly on 
for value of the energy $E_{n}$ in our numerical results.   

\begin{figure}[!t]
\vspfigA
\begin{center}
\includegraphics[width=\widthfig]{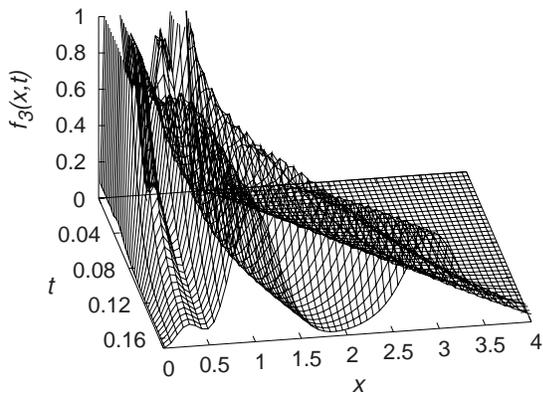}
\vspfigB
\caption{One-particle spatial distribution function 
   $f_{3}(x,t) \equiv \int_{0}^{+\infty} dx_{2}\; |\Psi_{3}(x,x_{2},t)|^{2}$ 
   as a function of time $t$ and position $x$, corresponding to the energy $E_{3}$, 
   for two particles with Coulomb interactions in a semi-infinite one-dimensional space $x\geq 0$. 
      Two peaks are seen in the time region $t>0.1$, corresponding to two particles.}
\label{fig3TwoSpaProb}
\end{center}
\vspfigC
\end{figure}  
%
   Figure \ref{fig3TwoSpaProb} is a graph of the one-particle spatial distribution function $f_{n}(x,t) \equiv \int_{0}^{+\infty} dx_{2}\; $ $|\Psi_{n}(x,x_{2},t)|^{2} = \int_{0}^{+\infty} dx_{1}\; |\Psi_{n}(x_{1},x,t)|^{2}$ for $n=3$ as a function of time $t$ and position $x$.  
   In this figure it is clearly seen that two peaks corresponding to two particles can survive for a long time although there are many peaks in a short time region. 
   The position of one peak of these two peaks closer to the wall at $x=0$ does not move very much in time, probably because the particle corresponding to this peak receives a repulsive Coulomb force from another particle to the direction of the wall and its escape movement is suppressed. 
   On the other hand, the position of another peak moves away quickly from the subspace.

\section{Chaos in Identical-Two-Particle Systems with Coulomb Interactions}
\label{EnergyLevelSpace}

   Now, we discuss a difference between one free particle and Coulomb-interacting two particles in a one-dimensional space in a viewpoint of quantum chaotic dynamics.    

   Chaos is defined dynamically as a classical system with a dynamical instability, i.e. a strong sensitivity to a small difference of initial conditions leading to at least one positive Lyapunov exponent, and quantum chaos has been interpreted as a quantum mechanical system whose classical counterpart is chaotic \cite{G90,O93,Sto99}. 
   In this sense, there is no quantum chaos in any one-dimensional one-particle system with any time-independent potential, because the corresponding classical system does not have the dynamical instability. 
   On the other hand, quantum two-particle systems in one-dimensional space can be chaotic, because the corresponding classical system can have a dynamical instability. 
   The quantum chaos has been also widely investigated by the distribution of energy level spacings \cite{G90,O93,Sto99}. 
   Following to this idea, the system with the time-reversible dynamics is regarded chaotic in a quantum sense if the distribution of energy level spacings is close to the Wigner distribution $(\pi/2) x \exp(-\pi x^{2}/4)$ by the Gaussian orthogonal ensemble (GOE), while it is non-chaotic if the distribution of energy level spacings is close to the Poisson distribution $\exp(-x)$. 
   Energy level spacing distributions of many-particle systems with Coulomb interactions have been calculated for two-particle cases in a one-dimensional space \cite{FST01}, in a one-dimensional space with a random potential \cite{SP01}, and in a two-dimensional space \cite{DSD04,STN09}, as well as for three-particle cases \cite{UP97}. 

\begin{figure}[!t]
\vspfigA
\begin{center}
\includegraphics[width=\widthfig]{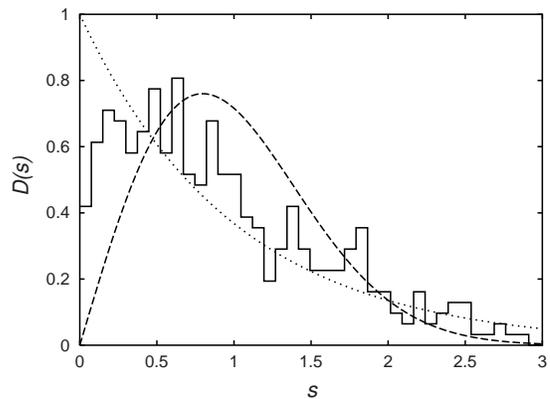}
\vspfigB
\caption{
   Distribution function $D(s)$ of energy level spacings (solid line) of two particles confined in the one-dimensional region $[0,l]$ without lead. 
   The broken line and the dotted line correspond to the Wigner distribution and the Poisson distribution, respectively. 
}
\label{fig4TwoLevSpa}
\end{center}
\vspfigC
\end{figure}  
%
   To calculate the energy level spacings of a quantum system we have to eliminate some trivial symmetric degrees of freedom from the Hamiltonian of the system. 
   The Hamiltonian of identical-two-particle systems with the Coulomb potential (\ref{Poten1}) is invariant for exchanging the two particle positions $x_{1}$ and $x_{2}$. 
   To eliminate the degeneracy caused by this symmetry from energy level spacings 
we calculated the energy eigenvalues of the spatially discretized Hamiltonian matrix (shown in Appendix \ref{SpaceDiscrite}) only for $x_{1}>x_{2}$. 
   (Here, we used the Fermion case so that by the Pauli exclusion principle the possibility of $x_{1}=x_{2}$ for the two particle positions $x_{1}$ and $x_{2}$ is zero.) 
   Further we also separate this Hamiltonian operator into the symmetric part and 
the anti-symmetric part for the transformation $x_{j}\rightarrow l - x_{j}, j=1,2$, 
and we take into account only of its symmetric part. 
   To calculate the distribution function of energy level spacings, we first calculate 
the energy eigenstates $\caE_{n}$, $n=1,2,\cdots,\mu$, $\caE_{1}\leq \caE_{2} \leq \cdots\leq\caE_{\mu}$ of this reduced Hamiltonian.  
   Then, we fit the indices $n=\mu_{0},\mu_{0}+1,\cdots,\mu$ as a function of energy $\caE_{n}$ by a smooth function $\rho (\caE) \equiv \sum_{k=1}^{10} a_{k} \caE^{k}$ with fitting parameters $a_{k}, k=1,2,\cdots,10$. 
   Here, we did not take into account the energy $\caE_{n}$ with small indices $n = 1,\cdots,\mu_{0}-1$ for this calculation with an integer $\mu_{0}>1$, in order to exclude strongly quantum regime from calculation of energy level spacings. 
   Using this fitting function we calculate the distribution function $D(s)$ of energy level spacings as the distribution function of $s \equiv \rho (\caE_{j+1}) - \rho(\caE_{j})$, $j=\mu_{0},\mu_{0}+1,\cdots,\mu-1$. 

   Figure \ref{fig4TwoLevSpa} is a graph of the distribution $D(s)$ of energy level spacings  (solid line) for a Coulomb-interacting identical-two-particle system confined in a one-dimensional region $[0,l]$ without lead.  
   Here, we used the same values of the system parameters $m$, $\hbar$, $l$, $\lambda$, $d$, and $\caN_{0}$ as used in Sec. \ref{CoulombInteractionEscape}, and also $\mu_{0} = 20$. 
   For comparisons, in Fig. \ref{fig4TwoLevSpa} we also show the Wigner distribution (broken line) and the Poisson distribution (dotted line).
   The distribution $D(s)$ in Fig. \ref{fig4TwoLevSpa} shows a repulsive behavior of energy level spacings and is different from the Poisson distribution, although its repulsion is not strong enough for the distribution $D(s)$ to be fit to the Wigner distribution. 
   In this sense, the identical-two-particle system with Coulomb interactions in a confined one-dimensional space is regarded to be weakly chaotic.

\section{Conclusion and Remarks} 

   In this paper we discussed escape behaviors of one- and two-particle systems from the one-dimensional finite region $(0,l)$ to the semi-infinite one-dimensional lead $[l,+\infty)$ with a positive constant $l$. 
   We prepared the initial condition of wave function of the system so that at the initial time $t=0$ 
the wave function is given by an energy eigenstate of the particle system confined in the finite region $(0,l)$. 
   Under these initial conditions we calculated the survival probability for particles to stay within this region $(0,l)$ at time $t$. 
   We showed that the survival probability decays in power $\sim t^{-3}$ asymptotically in one-dimensional one-free-particle systems, and it decays exponentially in two-particle systems with Coulomb interactions. 
   The quantum effect of identity of many particles such as the Pauli exclusion principle in a behavior of the survival probability were also discussed in the cases of Bosons and Fermions, and it was shown that the survival probability decays asymptotically in power $\sim t^{-6}$ for the identical two free Bosons but in power $\sim t^{-10}$ for the identical two free Fermions.

   This study is motivated to investigate, not only quantum and many-particle effects, 
but also a chaotic effect in escape phenomena. 
   In classical mechanical cases there is a conjecture by which the survival probability decays exponentially for chaotic systems while it decays in power for non-chaotic systems. 
   To check a chaotic feature of two-particle systems with Coulomb interactions we calculated the energy level spacing of the system in a confined one-dimensional space, and showed a repulsive feature of energy level spacings in this system. 
   This result suggests that two-particle systems with Coulomb interactions in a confined one-dimensional space are weekly chaotic. 
   On the other hand, one-free-particle systems in a one-dimensional space is not chaotic, because its corresponding classical systems do not have a dynamical instability.  
   Therefore, our results for quantum escapes in this paper may be consistent with the classical conjecture for a chaotic effect in exponential decays of survival probabilities. 

\begin{figure}[!t]
\vspfigA
\begin{center}
\includegraphics[width=\widthfig]{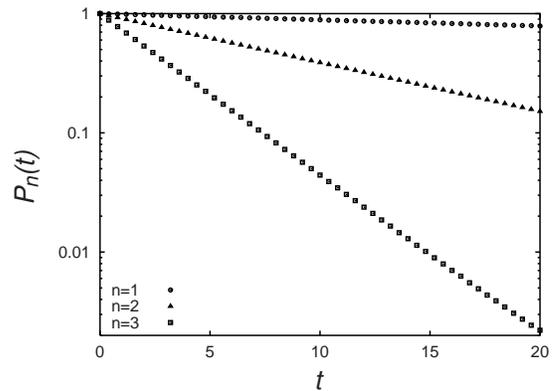}
\vspfigB
\caption{
   Survival probabilities $P_{n}(t), n=1,2,3$ of a one-particle system with a localized single impurity potential in a semi-infinite one-dimensional space as a function of time $t$ (linear-log plots), corresponding to the particle energies $E_{1}$, $E_{2}$, and $E_{3}$, respectively. 
   Here, the impurity is located just outside the subspace, and magnitude of the impurity potential is chosen to be very strong in comparison with the particle energies. 
   }
\label{fig5OneTunnel}
\end{center}
\vspfigC
\end{figure}  
%
   As a remark, although two-particle systems with Coulomb interactions show an exponential decay in the survival probability and they are (weekly) chaotic, it might not be necessary to mean that chaos is the origin of this exponential decay. 
   In other words, we still should be careful to a possibility that a chaotic feature is not a necessary condition for an exponential decay of survival probability in quantum systems.  
   As a result related to this point, in Fig. \ref{fig5OneTunnel} we show the survival probabilities $P_{n}(t), n=1,2,3$ for the one-particle system with a very localized single impurity at the just outside of subspace in a semi-infinite one-dimensional space, where we used parameter values as the subspace site number $\caN_{0}=60$, the total space site number $\caN = 16384$, and the discretized time interval $\delta t = 10^{-4}$. 
   Here, we chose the magnitude of the impurity potential as $1.8\times 10^{3}$, which is much larger than the particle energies $E_{1} = 4.77$ (circle), $E_{2}=  19.1$ (triangle), and $E_{3}= 42.9$ (square) in Fig. \ref{fig5OneTunnel}.  
   The one-particle system in a one-dimensional space is not chaotic, but Fig. \ref{fig5OneTunnel} shows a clear exponential decay in the survival probability of such a one-particle system with a strong impurity potential. 
   However, in Fig. \ref{fig5OneTunnel} the exponential decay rates of the survival probabilities for this one-particle case with an impurity strongly depend on the value of their particle energies, different from results of the two-particle systems with Coulomb-interactions as shown in Fig. \ref{fig3TwoPartiEsc}.

  It may also be meaningful to mention that an exponential decay of the survival probability in interacting two-particle cases shown in this paper occurs in a rather strong Coulomb interaction regime. 
  In cases with much smaller interaction magnitude $\lambda$ than used in Sec. \ref{CoulombInteractionEscape}, we observe a decay close to be in power, which is similar to the asymptotic decay behavior for identical two free particles shown in Sec. \ref{TwoFreeParticles}. 
  In a system with such a small $\lambda$, the distribution of energy level spacings is rather close to the Poisson distribution, implying that the system is rather close to be non-chaotic. 
  This may be a supportive evidence that a chaotic feature may be related to an exponential decay behavior of the survival probability at least in two-particle systems. 

   One may notice that in this paper exponential decays in two-particle systems with Coulomb interactions were shown numerically, so there is still a possibility that it might be only be a finite time property and it might decay in power in the long time limit $t\rightarrow +\infty$. 
   Related to this remark, Ref. \cite{DN02} argued that the survival probability for one-particle systems in an effectively one-dimensional space with a delta-functional impurity shows a power decay, even if it decays exponentially in time for a finite time region like in Fig. \ref{fig5OneTunnel}. 
   However, it would be valuable to show some results about differences between one-particle (non-chaotic) cases and two-particle (chaotic) cases even if they are finite time properties. 

   As a future problem on the subject of this paper it would be interesting to investigate escape behaviors of systems consisting of more than two particles, although in order to study such a large system numerically we would need much better numerical resources and techniques than used to obtain results in this paper. 
   Such large systems with interactions could be more strongly chaotic than two-particle systems, whose chaotic strength was not so strong as shown in Sec. \ref{EnergyLevelSpace}, and they could give better situations to study quantum chaotic effects in escape phenomena. 
   We could also consider a particle escape of such a large number of particle systems as a driving source to produce a particle current from a particle reservoir. 
   For such a case, as the initial state we could choose an equilibrium (e.g. canonical) state, which is represented as the density matrix produced by the energy eigenstates $\{\Phi_{n}(x_{1},x_{2},\cdots,x_{\tN})\}_{n}$ used in this paper with the weight of an equilibrium distribution.

\section*{Acknowledgements}

   One of the authors (T.T.) is grateful to Philippe A. Jacquet for stimulating discussions about escape problems. 
   This research was supported by the grant sponsor: The "Open Research Project for Physical Science of Biomolecular Systems" funded by the Ministry of Education, Culture, Sports, Science, and Technology of Japan.

  
\appendix

\section{Numerical Calculations of a One-Dimensional Two-Particle Dynamics}
\label{SpaceTimeDiscrite}

   In this Appendix we represent how we discretized the Schr\"odinger equation for the system consisting of two particles in a semi-infinite one-dimensional space to calculate the survival probability $P(t)$ in this paper. 
   Especially, we show a spatially discretized Hamiltonian and outline the pseudo-spectral method to solve the Schr\"odinger equation discretized in time.  

\subsection{Spatial Discretization of the Hamiltonian Operator}
\label{SpaceDiscrite}

   We consider two particles in a semi-infinite one-dimensional space and take $x_{j}(\geq 0)$ as the position coordinate of the $j$-th particle, $j=1,2$. 
   Then, we discretize the semi-infinite one-dimensional space by a positive constant $\dL$, so $x_{j} \rightarrow n_{j} \dL$, $n_{j} = 0, 1, 2, \cdots$ for $j=1,2$. 
   In this discretization of the one-dimensional space, 
the spatially second derivative $\partial^{2}/\partial x_{j}^{2}$ applying 
to any function $X(x_{j})$ is represented as 
\begin{eqnarray}
   \frac{\partial^{2} X(x_{j})}{\partial x_{j}^{2}}
      &\rightarrow& \frac{\tilde{X}(n_{j}+1)-2\tilde{X}(n_{j})+\tilde{X}(n_{j}-1)}{\dL^{2}} 
      \nonumber \\
    &\equiv& - \frac{1}{\dL^{2}}\sum_{k=0}^{+\infty} \caK_{n_{j}k}\tilde{X}(k)
\label{MatriKinet1}
\end{eqnarray}
where $\tilde{X}(n_{j}) \equiv X(n_{j} \dL)$ is the spatially discretized function of $X(x_{j})$. 
   Here, the matrix $\caK\equiv (\caK_{jk})$, $j=0,1,2,\cdots$, $k=0,1,2,\cdots$, is the matrix whose only nonzero elements are $\caK_{jj} = 2$ and $\caK_{(j+1)j} = \caK_{j(j+1)} = -1$, $j=0,1,2,\cdots$. 
   Using Eq. (\ref{MatriKinet1}), the Hamiltonian operator $\hat{H}$ is represented as the matrix $H$ for the spatially discretized representation: 
\begin{eqnarray}
    \hat{H} \rightarrow H \equiv \frac{\hbar^{2}}{2m\dL^{2}} K + \tilde{U}(\bfn) I
\label{HamilDiscr1}
\end{eqnarray}
with $\bfn \equiv (n_{1},n_{2})$ and $\tilde{U}(\bfn) \equiv U(n_{1}\dL,n_{2}\dL)$ using the potential $U(x_{1},x_{2})$ given by Eq. (\ref{Poten1}) for the continuous space case. 
Here, the matrices $K \equiv (K_{\bfn\bfn'})$ and $I \equiv (I_{\bfn\bfn'})$ are defined by 
\begin{eqnarray}
   K_{\bfn\bfn'} &\equiv& \caK_{n_{1}n_{1}'} \delta_{n_{2}^{}n_{2}'} 
      + \delta_{n_{1}^{}n_{1}'}  \caK_{n_{2}n_{2}'},  \\
   I_{\bfn\bfn'} &\equiv& \delta_{n_{1}^{}n_{1}'} \delta_{n_{2}^{}n_{2}'},  
\end{eqnarray}
respectively, for any $\bfn \equiv (n_{1},n_{2})$ and $\bfn' \equiv (n_{1}',n_{2}')$.  
Using Eq. (\ref{HamilDiscr1}) the Schr\"odinger equation is spatially discretized as 
$
   i\hbar \partial \tilde{\bfPsi}(t)/\partial t =  H \tilde{\bfPsi}(t)
$ 
as a equation for the vector $\tilde{\bfPsi}(t) \equiv (\tilde{\Phi}(\bfn,t))$ 
defined by $\tilde{\Phi}(\bfn,t) \equiv \Psi(n_{1}\dL,n_{2}\dL,t)$ with the vector index $\bfn$. 

   It may be meaningful to represent the Hamiltonian matrix $H$ by using the Dirac notation. 
   Introducing the state $\stateR{\bfn}$ as the one forming a complete ($\sum_{\bfn} \stateR{\bfn}\stateL{\bfn} = 1$) and orthogonal ($\left\langle\bfn|\bfn'\right\rangle = \delta_{n_{1}^{}n_{1}'} \delta_{n_{2}^{}n_{2}'}$) set for the site indexes $\bfn$ and $\bfn'$, the Hamiltonian matrix $H$ can be represented as the operator 
%
\begin{eqnarray}
   \hat{\caH} \equiv \sum_{\bfn} \stateR{\bfn}\epsilon_{\bfn}\stateL{\bfn} 
   + u \sum\limits_{\sumindex{\bfn,\bfn'}{|\bfn-\bfn'|=1}} \stateR{\bfn}\stateL{\bfn'} 
\label{HamilTight1}
\end{eqnarray}
where $\epsilon_{\bfn}$ and $u$ are defined by 
$
   \epsilon_{\bfn} \equiv [2\hbar^{2}/(m\dL^{2})] 
      + U(n_{1}\dL,n_{2}\dL)
$ and $u \equiv - \hbar^{2}/(2m\dL^{2})$, respectively. 
   The operator (\ref{HamilTight1}) has the same type of form as a tight-binding Hamiltonian with the site energy $\epsilon_{\bfn}$ and the hopping rate $u$, and the Schr\"odinger equation is represented as 
$ 
   i\hbar \partial \stateR{\Psi(t)}/\partial t = \hat{\caH}  \stateR{\Psi(t)}
$
as a equation for the state $\stateR{\Psi(t)} \equiv \sum_{\bfn}\tilde{\Phi}(\bfn,t) \stateR{\bfn}$.

\subsection{Time-Discretization of Schr\"odinger Equation by Pseudo-Spectral Method} 
\label{PseudoSpectralMethod}

   In the previous subsection of this Appendix we discussed how we spatially discretized the Hamiltonian operator. 
   In this subsection we outline how we discretize the time-evolution by the Schr\"odinger equation in the way called by the pseudo-spectral method.    

   We consider a one-dimensional space of the length $L$ consisting of the subspace and the lead, and note that the function $\caX(x_{1},x_{2})$ of $x_{1}$ and $x_{2}$ satisfying the boundary condition $\caX(0,x_{2}) = \caX(x_{1},0) =\caX(L,x_{2}) = \caX(x_{1},L) = 0$ can be Fourier-transformed as 
\begin{eqnarray}
   \tilde{\caX}(k_{1},k_{2}) &=& \sqrt{\frac{2}{L}}\int_{0}^{L}dx_{1}\int_{0}^{L}dx_{2}\;  
      \caX(x_{1},x_{2}) 
      \nonumber \\
   &&\spaEq\times
   \sin\!\left(\frac{\pi k_{1}}{L} x_{1}\right)\sin\!\left(\frac{\pi k_{2}}{L} x_{2}\right)
      \nonumber \\
   &\equiv& \hat{\caF} \left[\caX(x_{1},x_{2})\right] ,
     \\
   \caX(x_{1},x_{2}) &=& \sqrt{\frac{2}{L}}\sum_{k_{1}=1}^{+\infty}\sum_{k_{2}=1}^{+\infty} 
      \tilde{\caX}(k_{1},k_{2}) 
      \nonumber \\
   &&\spaEq\times
   \sin\!\left(\frac{\pi k_{1}}{L} x_{1}\right)\sin\!\left(\frac{\pi k_{2}}{L} x_{2}\right)
      \nonumber \\
   &\equiv& \hat{\caF}^{-1} \left[ \tilde{\caX}(k_{1},k_{2})\right]
      \label{FInv1}
\end{eqnarray}
by using the relation $\int_{0}^{L} dx \; \sin (\pi k x/L)\sin (\pi k' x/L) = L\delta_{kk'}/2$, $k=1,2,\cdots$, $k'=1,2,\cdots$  etc. 
   Using Eq. (\ref{FInv1}) we obtain 
\begin{eqnarray}
   \hat{K} \caX(x_{1},x_{2}) = \hat{\caF}^{-1} \left[ \tilde{K}(k_{1},k_{2}) \tilde{\caX}(k_{1},k_{2})\right]
\label{KinetOpera1}
\end{eqnarray}
with the kinetic operator $\hat{K}\equiv -[1/(2m)](\partial^{2}/\partial x_{1}^{2}+\partial^{2}/\partial x_{2}^{2})$ and the function $\tilde{K}(k_{1},k_{2}) \equiv [\pi^{2}/(2mL^{2})]\left(k_{1}^{2}+k_{2}^{2}\right)$ of $k_{1}$ and $k_{2}$. 

   We discretize the time by a positive constant $\delta t$, so $t \rightarrow \nu \delta t, \nu = 0,1,2,\cdots$. 
   By using the formal solutions of the Schr\"odinger equation, the wave function $\Psi(x_{1},x_{2},t+\delta t)$ at time $t+\delta t$ is related to the wave function  $\Psi(x_{1},x_{2},t)$ at time $t$ as 
\begin{widetext}
\begin{eqnarray}
   \Psi(x_{1},x_{2},t+\delta t) 
      &=& e^{-i\hat{H}\delta t/\hbar}\Psi(x_{1},x_{2},t)
      \label{TimeEvolu0} \\
   &=& e^{-iU(x_{1},x_{2})\delta t/(2\hbar)}e^{-i\delta t\hat{K}/\hbar}
      e^{-iU(x_{1},x_{2})\delta t/(2\hbar)}
      \Psi(x_{1},x_{2},t) +\mathcal{O}(\delta t^{2})
      \nonumber \\
   &=& e^{-iU(x_{1},x_{2})\delta t/(2\hbar)}\hat{\caF}^{-1} 
      \left[e^{-i\delta t \tilde{K}(k_{1},k_{2}) /\hbar}
       \hat{\caF} \left[ e^{-iU(x_{1},x_{2})\delta t/(2\hbar)}
     \Psi(x_{1},x_{2},t) \right]\right] +\mathcal{O}(\delta t^{2}) 
\label{TimeEvolu1}
\end{eqnarray}
\end{widetext}
where we used the relation (\ref{KinetOpera1}) and the boundary conditions  $\Psi(0,x_{2},t) = \Psi(x_{1},0,t) =\Psi(L,x_{2},t) = \Psi(x_{1},L,t) = 0$ for the wave function of the system used in this paper.
   By Eq. (\ref{TimeEvolu1}) we can calculate the wave function $\Psi(x_{1},x_{2},t+\delta t)$ at time $t+\delta t$ from the wave function $\Psi(x_{1},x_{2},t)$ at the previous time $t$. 

   An advantage of the pseudo-spectral method  (\ref{TimeEvolu1}) is that in this method we do not have to apply the space-differential operator $e^{-i\hat{H}\delta t/\hbar}$ in the time-evolution of wave function, and it is replaced by simple multiplications of just the numbers $ e^{-iU(x_{1},x_{2})\delta t/(2\hbar)}$ and $e^{-i\delta t \tilde{K}(k_{1},k_{2}) /\hbar}$, leading to less complicate numerical calculations than to use Eq. (\ref{TimeEvolu0}) directly.
   Instead, we need to do a Fourier transformation and an inverse Fourier transformation for one step of the time evolution, but we can use the technique called by the fast Fourier transformation \cite{D94} in actual numerical calculations. 
   The fast Fourier transformation requires the calculation time proportional to be $\tilde{\caN} \log_{2}\tilde{\caN}$ (instead of $\tilde{\caN}^{2}$) for the total ($x_{1}x_{2}$ space) site number $\tilde{\caN}$, and it is a big advantage for a fast numerical calculation of large $\tilde{\caN}$ systems such as used in this paper. 
   

\section{Survival Probability of Free Particle Systems in the semi-infinite one-dimensional space}

   In this appendix we calculate the survival probability analytically for free particle systems 
in a semi-infinite one-dimensional space.  
   First, we calculate it for the case of one free particle, and 
show analytically an asymptotic power decay (\ref{EscapRateFree1}) of the survival probability $P(t)$. 
   Secondly, we show the different power decay behaviors (\ref{EscFreeTwoBoson1}) and (\ref{EscFreeTwoFermi1}) of the survival probabilities between identical two free Bosons and Fermions with the quantum effect of identity of two particles, such as the Pauli exclusion principle for Fermions,  in a semi-infinite one-dimensional space.

\subsection{One Free Particle Case}
\label{EscapeFreeParticle}

   For one free particle in a one-dimensional space, the Hamiltonian operator is given by $\hat{H} = -[\hbar^{2}/(2m)] \partial^{2}/\partial x^{2}$. 
   Then, the wave function $\Psi^{'}(x,t)$ for this Hamiltonian system in the full one-dimensional infinite region $(-\infty,+\infty)$ is represented as  \cite{FH65}
\begin{eqnarray}
   \Psi^{'}(x,t) &=& \sqrt{\frac{m}{2\pi i\hbar t}} \int_{-\infty}^{+\infty}dy\; \Psi^{'}(y,0) 
      \nonumber \\
    &&\spaEq\spaEq\spaEq\spaEq \times 
       \exp\!\left[\frac{im(x-y)^{2}}{2\hbar t}\right] 
\end{eqnarray}
for any initial wave function $\Psi^{'}(x,0)$ of the system.  
   Using this function $\Psi^{'}(x,t)$, the wave function $\Psi(x,t)$ for the system with the same Hamiltonian but in the semi-infinite one-dimensional region $[0,+\infty)$ is given by 
\begin{eqnarray}
   \Psi(x,t) &=& \Xi^{'}(t)^{-1}[\Psi^{'}(x,t) - \Psi^{'}(-x,t)] 
      \label{WaveFunctFree0a} \\
   &=& \sqrt{\frac{m}{2\pi i\hbar t}} \int_{-\infty}^{+\infty}dy\; 
      \Psi(y,0) 
      \nonumber \\
   &&\spaEq\spaEq\spaEq\spaEq \times 
      \exp\!\left[\frac{im(x-y)^{2}}{2\hbar t}\right]
\label{WaveFunctFree0b}
\end{eqnarray}
for $x\geq 0$, so that the boundary condition $\Psi(0,t) = 0$ in the hard-wall at $x=0$ is automatically satisfied at any time $t$ by Eq. (\ref{WaveFunctFree0a}). 
   Here,  $\Xi^{'}(t)\equiv\int_{0}^{+\infty}dx\;|\Psi^{'}(x,t) - \Psi^{'}(-x,t)|^{2}$ is the quantity to normalize the wave function $\Psi(x,t)$ as $\int_{0}^{+\infty} dx\; |\Psi(x,t)|^{2} = 1$ for the semi-infinite space $[0,+\infty)$, and the initial wave function $\Psi(x,0) = \Xi^{'}(0)^{-1}[\Psi^{'}(x,0) - \Psi^{'}(-x,0)]$ satisfies the condition  $\Psi(-x,0) = -\Psi(x,0)$ for any real number $x$ in Eq. (\ref{WaveFunctFree0b}).
   By using Eq. (\ref{WaveFunctFree0b}) and noting the fact that value of the initial wave function $\Psi(x,0)$ is zero for $|x| > l$, the wave function $\Psi(x,t)$ of one free particle in the semi-infinite region $[0,+\infty)$ is represented as  
\begin{eqnarray}
   \Psi(x,t) = \int_{0}^{l}dy\; G(x,y;t) \Psi(y,0)
\label{WaveFunctFree1}
\end{eqnarray}
at the position $x$ at time $t$ with $G(x,y;t)$ defined by 
\begin{eqnarray}
   G(x,y;t) &=& \sqrt{\frac{m}{2\pi i\hbar t}} \left\{\exp\!\left[\frac{im(x-y)^{2}}{2\hbar t}\right]
      \right.\nonumber \\
   &&\spaEq\spaEq\spaEq\left. 
      - \exp\!\left[\frac{im(x+y)^{2}}{2\hbar t}\right]\right\} , \;\;\;\;\;\;
\label{PropaOne1}
\end{eqnarray}
as the time-evolutional propagator for one free particle in the semi-infinite one-dimensional space $[0,+\infty)$. 

   For large $t$ we expand the propagator $G(x,y;t)$ as 
\begin{eqnarray}
   G(x,y;t) &=& -\sqrt{\frac{2i}{\pi}}\left(\frac{m}{\hbar t}\right)^{3/2}xy
      +\mathcal{O}\!\left(t^{-5/2}\right) 
\label{Kexpan1}
\end{eqnarray}
up to the smallest non-zero order of $1/t$. 
   Inserting Eq.  (\ref{Kexpan1}) into Eq. (\ref{WaveFunctFree1}) and then calculating  the survival probability (\ref{SurviProba1}) for $\tN=1$, we obtain Eq. (\ref{EscapRateFree1}) with the coefficient (\ref{ContA1}).

\subsection{Identical Two Free Particle Cases}

   We consider an identical-two-free-particle system with no potential energy in a semi-infinite 
one-dimensional space. 
   In this case, because of no potential energy, the time-evolution of the wave function 
$\Psi(x_{1},x_{2},t)$ of this system is dominated by the one-particle propagator (\ref{PropaOne1}) 
and is given by 
\begin{eqnarray}
   \Psi(x_{1},x_{2},t) &=& \int_{0}^{l}dy_{1}\int_{0}^{l}dy_{2}\; 
      G(x_{1},y_{1};t) G(x_{2},y_{2};t) 
      \nonumber \\
   &&\spaEq\spaEq\times
      \Psi(y_{1},y_{2},0) .
\label{WaveFunctTwo2}
\end{eqnarray}
Here, we used the fact that values of the wave function $\Psi(y_{1},y_{2},0)$ at the initial time $t=0$ are nonzero only in the region satisfying $0<y_{1}<l$ and $0<y_{2}<l$ as assumed in this paper. 

   Now, we impose the condition 
\begin{eqnarray}
   \Psi(x_{2},x_{1},0) = \pm \Psi(x_{1},x_{2},0),  
\label{ExchaInterIniti2}
\end{eqnarray}
i.e. the condition (\ref{ExchaInter1}) at the initial time $t=0$. 
   Here, the sign $+$ ($-$) in the right-hand side of Eq. (\ref{ExchaInterIniti2}) is taken for Bosons (Fermions). 
   Under the condition (\ref{ExchaInterIniti2}) the wave function (\ref{WaveFunctTwo2}) automatically satisfies the condition (\ref{ExchaInter1}) at any time $t$. 
   We can rewrite Eq. (\ref{WaveFunctTwo2}) using Eq. (\ref{ExchaInterIniti2}) as 
\begin{eqnarray}
   &&\Psi(x_{1},x_{2},t) = \frac{1}{2}\left[\Psi(x_{1},x_{2},t)\pm\Psi(x_{2},x_{1},t)\right]
      \nonumber \\
   &&\;\;\; 
      = \frac{1}{2}\int_{0}^{l}dy_{1}\int_{0}^{l} dy_{2}\; \left[
      G(x_{1},y_{1};t) G(x_{2},y_{2};t) 
      \right.\nonumber \\
   && \spaEq\spaEq \left.\pm  G(x_{2},y_{1};t) G(x_{1},y_{2};t)\right]\Psi(y_{1},y_{2},0) .\spaEq
\label{WaveFunctTwo3}
\end{eqnarray}
so that the wave function $\Psi(x_{1},x_{2},t)$ automatically satisfies the condition (\ref{ExchaInter1}) at any time $t$ without noting the condition  (\ref{ExchaInterIniti2}) anymore.

\subsubsection{Boson Case}
\label{TwoFreeBosons}

For the system consisting of identical two free Bosons, by Eq. (\ref{Kexpan1}) the quantity $G(x_{1},y_{1};t) G(x_{2},y_{2};t) +  G(x_{2},y_{1};t) G(x_{1},y_{2};t)$ is asymptotically represented as
\begin{eqnarray}
   &&G(x_{1},y_{1};t) G(x_{2},y_{2};t) +  G(x_{2},y_{1};t) G(x_{1},y_{2};t) 
      \nonumber \\
   &&\;\;\; = 
      \frac{4i}{\pi}\left(\frac{m}{\hbar t}\right)^{3}x_{1}x_{2}y_{1}y_{2}
      +\mathcal{O}\!\left(t^{-4}\right) .
\label{PropaTwo2}
\end{eqnarray}
   Inserting Eq.  (\ref{PropaTwo2}) into Eq. (\ref{WaveFunctTwo3}) and then 
calculating the survival probability (\ref{SurviProba1}) for $\tN=2$, we obtain Eq. (\ref{EscFreeTwoBoson1}) with the coefficient (\ref{ConstBoson1}). 

\subsubsection{Fermion Case}
\label{TwoFreeFermions}

   For the system consisting of identical two free Fermions we expand the one-particle propagator (\ref{PropaOne1}) as 
\begin{eqnarray}
   && G(x,y;t) 
      \nonumber \\
   &&\;\;\; = \sqrt{\frac{2i}{\pi}}\left(\frac{m}{\hbar t}\right)^{3/2}xy
      \left[1+\frac{im}{2\hbar t}\left(x^{2}+y^{2}\right)
      \right.\nonumber \\
   && \spaEq \left.
      -\frac{1}{24}\left(\frac{m}{\hbar t}\right)^{2}\left(3x^{2}+10x^{2}y^{2}+3y^{2}\right)
      +\mathcal{O}\!\left(t^{-3}\right) \right] 
     \spaEq\nonumber \\
\label{Kexpan2}
\end{eqnarray}
up to the order of $1/t^{5/2}$, which is higher order than in Eq. (\ref{Kexpan1}). 
   By Eq. (\ref{Kexpan2}) the quantity $G(x_{1},y_{1};t) G(x_{2},y_{2};t) -  G(x_{2},y_{1};t) G(x_{1},y_{2};t)$ is asymptotically represented as
\begin{eqnarray}
   &&G(x_{1},y_{1};t) G(x_{2},y_{2};t) -  G(x_{2},y_{1};t) G(x_{1},y_{2};t) 
      \nonumber \\
   &&\;\;\; = \frac{1}{3\pi i}\left(\frac{m}{\hbar t}\right)^{5} 
      x_{1}x_{2}y_{1}y_{2}\left(x_{1}^{2}-x_{2}^{2}\right)\left(y_{1}^{2}-y_{2}^{2}\right)
      \nonumber \\
   &&\spaEq
      +\mathcal{O}\!\left(t^{-6}\right) .
\label{PropaTwo3}
\end{eqnarray}
   Inserting Eq.  (\ref{PropaTwo3}) into Eq. (\ref{WaveFunctTwo3}) and then 
calculating the survival probability (\ref{SurviProba1}) for $\tN=2$, we obtain Eq. (\ref{EscFreeTwoFermi1}) with the coefficient (\ref{ContFermi1}).



\end{document}